# Identification of complex systems in the basis of wavelets

*Alexander Shaydurov*
*McGill University, RRP*
*shaydurov@ieee.org*

**Abstract**
In this paper is proposed the method of the identification of complex dynamic systems. Method can be used for the identification of linear and nonlinear complex dynamic systems for the determined or stochastic signals at the inputs and the outputs. It is proposed to use a basis of wavelets for obtaining the impulse transient function (**ITF**) of system. **ITF** is considered in the form of surface in the **3D** space. Are given the results of experiments on the identification of systems in the basis of wavelets.

**Introduction**
Under the dynamic system $\Sigma$ will be understood the ordered set of the elements of any nature: energy, social, economic, information and any others, united by connections, that represent integral unity and having by memory [2,5,10]. The set-theoretical model of complex dynamic system can be represented:

$$\sum = (T, R, \Omega, Y, \varphi, \eta)$$

where **T** - set of moments of time, which fix the states of system **R**; $\Omega$ - many input controlled ($\Omega_1$) and un controlled ($\Omega_2$) signals of the system:

$$\Omega_1 = \{\omega_1 : T \to X_1\};$$
$$\Omega_2 = \{\omega_2 : X_2\};$$
$$X = X_1 \cup X_2,$$

where $\omega_1, \omega_2$ - the defined probability measures; $\varphi$ - the operator of the passages of the system:

$$\varphi : T \times T \times R \times \{(t, X_L]_{t_0}^t\} \to R$$

$(t, X_L]_{t_0}^t$ - the fragment of input communication; $\eta$ - the operator of the outputs of the system:

$$\eta(t, t_0, R(t_0), \{(t, X_L]_{t_0}^t\} \to Y,$$

Set of outputs of system **Y** are determined by the collection of input signals and by certain of generalized operator, in the general case nonlinear, assigning mapping the set of input signals onto set output (operator of the approximation of function the output signals of the system through the function of input signals). As this operator it is convenient to use **ITF** of system - **h(t)** [3,6,9]. In the general case input signals are nonstationary stochastic processes.

Let us define the task of the identification of system as determination **ITF**, which unites inputs ($\varphi$) and outputs operators ($\eta$) and directly reflecting the inputs of system **X** into the set outputs **Y**. Takeing into account many possible versions of systems, it is utilized the classification of systems in the form [6] $A = <\alpha\beta\gamma\delta>$, where $\alpha$ - the sign of the dynamicity of system (0,1): $\beta$ - stochasticity of processes; $\gamma$ - the sign of nonlinearity; $\delta$ - discretion. Output for the dynamic determined linear systems ($A = <100\delta>$) is determined by equation the convolution:

$$Y(t) = \int_0^t h(\tau) X(t-\tau) d\tau, \quad (1)$$

and for the dynamic stochastic linear systems ($A = <110\delta>$) - by Wiener-Hopf equation:

$$R_{xy}(t) = \int_0^T h(\tau) R_{xx}(t-\tau) d\tau, \quad (2)$$

where $R_{xy}(t)$, $R_{xx}(\tau)$ with respect the cross- and autocorrelation functions of input and output. For the systems of class $A = <1\beta 1\delta>$ (dynamic determined or stochastic nonlinear) adapt the equations of Volterra, sharply increasing size of calculations, model of Wiener or Hammerstein, which select from the system linear and nonlinear parts and their characterized by sequence arrangements [2,3,7].

The appearance of nonlinear properties, the need for the calculation of stochastic and nonstationary processes, very probable and frequently meeting with a real systems, make the task of the identification of extremely complex both with the systematic and the computational of the points of view. The use of Fourier transforms under these conditions and the solutions of equations (1) and (2) in the frequency domain become incorrect [10].

The very promising trend of the permission of the emergent problem is use wavelet's analysis [1,4]. The generation of the additional frequencies, which disrupt the principle of superposition, is the main sign of the nonlinearity of system. With the use wavelet's transform the system it can be analyzed with the wide frequency spectrum, and each separate frequency range can be considered as "frequency channel." The isolation of the frequency channels of input and output it makes it possible to count the system of linear for each channel.

Since wavelet's transform it is linear, then passages from the region of real signals into the region wavelet coefficients and vice versa they do not introduce any distortions. To carry out the necessary conversions is possible in any most convenient region.

*Method identification*

With the use of continuous wavelet transform the equation of relation of input and output of linear dynamic system can be presented:

$$W^Y[a,b] = W[\int_0^T h(\tau) X(t-\tau) d\tau], \quad (3)$$

and the equations of convolution to write down in the form:

$$W\left\{\int_{-\infty}^{+\infty} f_1(t-\tau) f_2(\tau) d\tau\right\} =$$

$$= \frac{1}{\sqrt{|a|}} \int_{-\infty}^{+\infty} \{\int_0^T f_1(t-\tau) f_2(\tau) d\tau\} \psi(\frac{t-b}{a}) dt$$

or, converting [ 1 ]:

$$\frac{1}{\sqrt{|a|}} \int_{-\infty}^{+\infty} f_2(\tau) \{\int_0^T f_1(t-\tau) \psi(\frac{t-b}{a}) dt\} d\tau =$$

$$= \int_0^T f_2(\tau) \{\frac{1}{\sqrt{|a|}} \int_{-\infty}^{+\infty} f_1(t-\tau) \psi(\frac{t-b}{a}) dt\} d\tau .$$

It wavelet coefficients we obtain on the basis of the property of displacement:

$$\frac{1}{\sqrt{|a|}} \int_{-\infty}^{+\infty} f_1(t-\tau) \psi(\frac{t-b}{a}) dt = W[a, b-\tau]$$

Then

$$W^Y[a,b] = \int_0^T f_2(\tau) W^X[a, b-\tau] d\tau$$

or, using ITF - **h(τ)**:

$$W^Y[a,b] = \int_0^T h(\tau) W^X[a, b-\tau] d\tau$$

Let, $a = a^* - const$ i.e., be examined the specific frequency channel:

$$W^Y[a^*, b] = \int_0^T h(\tau) W^X[a^*, b-\tau] d\tau , \quad (4)$$

where, $W^Y[a^*, b]$, $W^X[a^*, b-\tau]$ they are functions one by the variable **b**, which reflects time. Formula (4) is equation of convolution:

$$F_{a^*}^Y(b) = \int_0^T h(\tau) F_{a^*}^X(b-\tau) d\tau ,$$

where, $F_{a^*}^Y(b) = W^Y[a,b]$,

$F_{a^*}^X(b-\tau) = W^W[a^*, b-\tau]$ the section of surface it wavelet coefficients with $a = a^*$.

Fourier transform for each definite frequency channel can be recorded in the form:

$$F_{a^*}^Y(\omega) = H_{a^*}(\omega) F_{a^*}^X(\omega), \quad (5)$$

From where the frequency image of the section of surface ITF with $a = a^*$.

$$H_{a^*}(\omega) = \frac{F_{a^*}^Y(\omega)}{F_{a^*}^X(\omega)},$$

The reverse Fourier transform $H(\omega) \to h(t)$ makes possible to pass in for the image of surface ITF.

Thus ITF for the complex system $A = <\alpha\beta\gamma\delta>$ it is defined as surface in the space of time- frequency coordinates.

Using the developed mathematical method it was created software of analysis and identification of complex systems, that includes wavelet the analysis:

discrete (Daubechies wavelets) and continuous (DOG, MHAT, MORLET, SHAHHON, PAUL, GAUSS and other.) it wavelet transform; the statistical analysis of the systems: the determination of the statistical parameters, auto- and cross correlation dependence, the detection of the distribution laws, the spectral analysis of functions, the generation of random functions according to the assigned distribution laws.

**Experimental results**

Simulation starting asynchronous machine

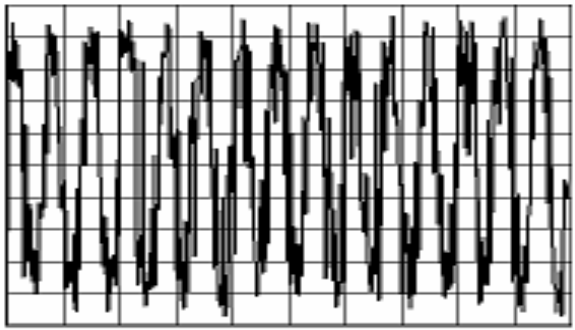

Figure 1. Input signal – supply voltage (stochastic processes)

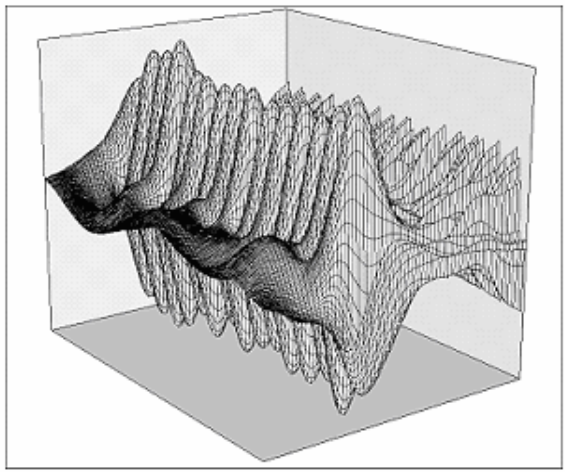

Figure 2. Surface of wavelet's coefficients of input signal.

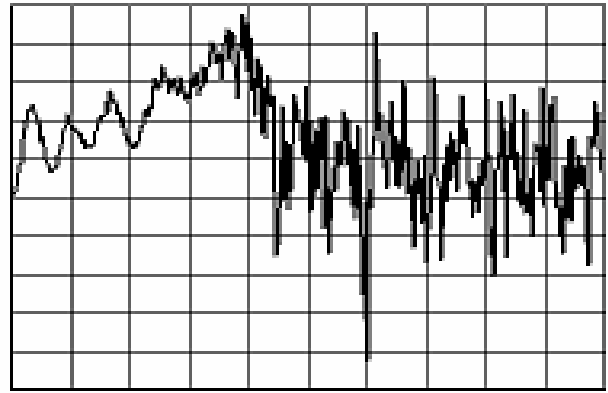

Figure 3. Output signal – electromagnetic torque

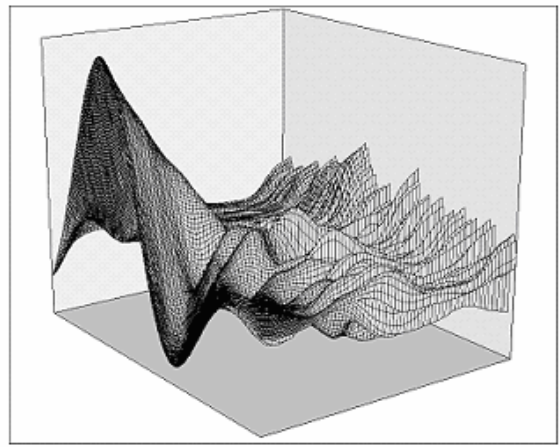

Figure 4. Surface of wavelet's coefficients of output signal.

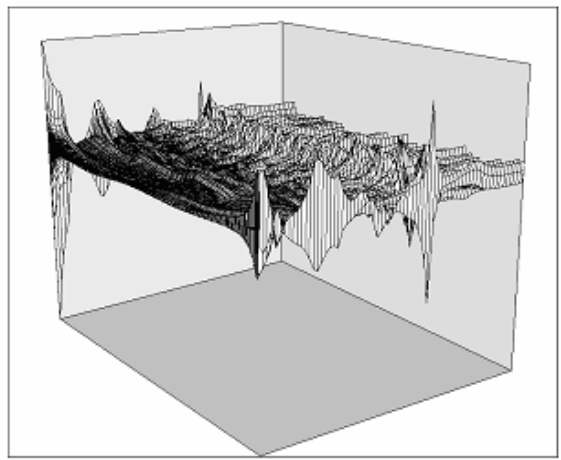

Figure 5. Surface of wavelet's coefficients of ITF.

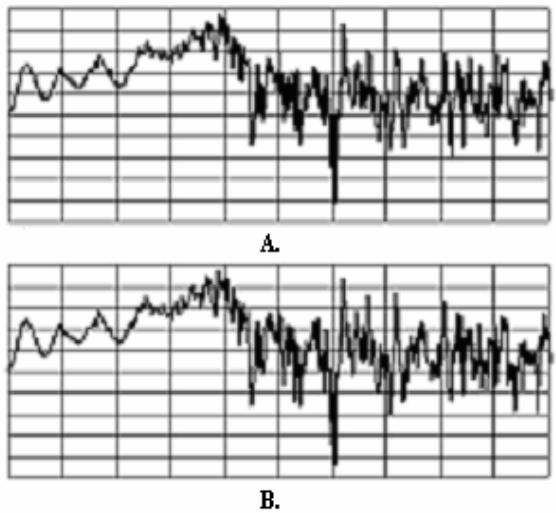

Figure 6. Original output signal (**A**) and restore (**B**) signal. Error of restore ε = 1.311

**Conclusions**

The new method of study and identification of complex nonlinear dynamic systems is proposed. It possesses high efficiency and speed. Software are developed and computational experiments with the nonlinear systems, which have nonstationary, stochastic signals at the inputs and the outputs carrying out. Experiments showed high accuracy of the restoration of signals at the outputs on the known input and ITF of system. The proposed method can be successfully used for the tasks of the approximation of functions by functions, i.e., the tasks, solved at present by neuron networks. In the telecommunication this method can be used for determining the property of the information channel with the known input and the permissible distortions of output.

The use of the obtained surface ITF makes it possible to solve the problems of the synthesis of the models of complex systems in the form of electromagnetic or other analogs

**References**

[1]. Doubechies. Ten Lectures on Wavelets. CBMS-NSF Regional Conf. Ser. In Appl. Math. Vol.61, SIAM, Philadelphia, 1992.

[2]. Eyknoff P. System identification. Parametr and state estimation. University of Technjlogy Eindhoven, trhe Netherlands, 1975.

[3]. E. Ikonen. Advanced Process Identification and Control. Marcel Dehher Inc. NY, 2002.

[4]. F.Koinert Wavelets and Multiwavelets. Chapman & Hall/CRC,2004.

[5]. V. Lin General Systems Theory. A Mathematical Approach. Kluwer Academic Publishers, NY,2002.

[6]. Rastrigin L.A., Madgerov N.E. Introduction to identfication of control objects. Moscow. Energy, 1977.

[7]. W.Rugh. Nonlinear Systems Theory. The Volterra/Wiener Approach. The Johns Hopkina. University Press, 1981.

[8]. E.Scheinerman Invitation to Dynamic Systems. Prence Hall, New Jersey, 2001.

[9]. Shaydurov A., 2004, "Identification and synthesis of the models electromechanical energy converters", CCECE 2004, Canadian conference on Electrical and Computer Engineering, ISBN 0-7803-8254-4.

[10]. Shaydurov A., 1993, "Development of the theory of systems analysis and construction of the optimum Power Electrical Systems", Doc. Eng. Sc. Thesis, Moscow.